\documentstyle[prl,aps,epsf]{revtex}

\begin{document}

\twocolumn[\hsize\textwidth\columnwidth\hsize\csname%
@twocolumnfalse\endcsname%
 
\title{Order parameter of quasi-one-dimensional superconductors: \\
symmetry features in quasiparticle density of 
states and spin susceptibility}

\author{R. D. Duncan, C. D. Vaccarella, and C. A. R. S\'a de Melo
\\ School of Physics, Georgia Institute of Technology \\
Atlanta, GA 30332}

%\pacs{74.70.Kn}

\date{February 23, 2001}
\maketitle

\begin{abstract}
Recent experiments indicate that the Bechgaard Salts
$\rm (TMTSF)_2 ClO_4$ and ${\rm (TMTSF)_2 PF_6}$ may be unconventional
triplet superconductors.
The quasiparticle density of states and the uniform spin susceptibility
tensor are computed at low temperatures for 
order parameter symmetries, as an attempt to narrow the number of  
possibilities based on current experimental evidence. 
\vskip 0.3cm
\end{abstract} 
]

%%\section{Introduction}
%%\label{introduction}

In early theoretical investigations of 
quasi-one-dimensional superconductors
(of the Bechgaard family $ {\rm (TMTSF)_2 ClO_4} $ and 
${ \rm (TMTSF)_2 PF_6 } $) in the presence of an external magnetic field, 
the possibility of triplet behavior 
in these systems was suggested~\cite{lebed-86,dms-93,sdm-96}.
The present experimental evidence indicates 
that these quasi-one-dimensional superconductors are indeed 
unconventional~\cite{lee-94,lee-95,lee-97,lee-00}.
For instance, Lee {\it et. al.}~\cite{lee-97} 
measured the magnetic field versus temperature phase diagram 
for ${ \rm (TMTSF)_2 PF_6}$ under pressure of 6 kbar.
They found that the upper critical fields along the usual 
${\bf a}$, ${\bf b^{\prime}}$,
and ${\bf c}$ direction are highly anisotropic.
Furthermore, the Pauli paramagnetic limit is exceeded by a factor
of 3 or 4.
In addition, Lee {\it et. al.}~\cite{lee-00} found that
there is no Knight shift in ${ \rm (TMTSF)_2 PF_6 } $ 
for fields ${\bf H} \parallel {\hat b}^{\prime}$
at pressures $P \approx 6~{\rm kbar}$.
This suggests the existence of a triplet superconducting
phase in this system. The temperature versus magnetic field 
phase diagram of $ {\rm (TMTSF)_2 ClO_4} $ at ambient pressure
was also measured by Lee {\it et. al.}~{\cite{lee-94,lee-95}, for 
${\bf H} \parallel {\hat b}^{\prime}$. 
These measurements seem to indicate
that the Pauli paramagnetic limit is exceeded in this compound,
which suggests triplet superconductivity.
Furthermore, Belin and Behnia~\cite{behnia-97} reported 
measurements of the thermal conductivity in the superconducting state of 
$ {\rm (TMTSF)_2 ClO_4} $, indicating their data is inconsistent 
with the existence of gap nodes at the Fermi surface. 
These experimental results combined suggest the existence
of a fully gapped triplet superconducting state 
in ${\rm (TMTSF)_2 ClO_4}$.
Although the similar crystal structure
of these systems suggests, from a simple group theoretical point of view, 
that the origin of the pair interaction is the same, the role of the
proximity to an SDW phase in $ {\rm (TMTSF)_2 PF_6} $, 
for instance, 
needs to be investigated both theoretically and experimentally. 
Furthermore, the order parameter symmetry also needs to be 
investigated via phase sensitive experiments like those 
performed in cuprate oxides~\cite{harlingen-93,wellstood-95}.

In this paper we are concerned with the symmetry of the order parameter 
of a triplet quasi-one-dimensional superconductor at zero magnetic field
and we have three main goals.
First, we perform a group theoretical analysis of the possible symmetries
of the order parameter for an orthorhombic quasi-one-dimensional 
superconductor at zero magnetic field. Second, we calculate
the quasiparticle density of states at zero temperature and
the uniform spin susceptibility tensor at low temperatures for various
candidate symmetries of the order parameter consistent with our
group theoretical analysis. Third, we make connections to 
scanning tunneling microscopy (STM) 
of quasiparticle density of states and Knight shift measurements of 
the spin susceptibility tensor.

%%\section{Hamiltonian}
%%\label{sec:Hamiltonian}

We study quasi-one-dimensional 
systems with a single band, in an orthorhombic
lattice, and allow for singlet or triplet pairing. 
We consider the following dispersion 
\begin{equation}
\label{eqn:dispersion}
\epsilon_{\bf k} = -|t_x| \cos(k_x a) - |t_y| \cos(k_y b) 
- |t_z| \cos(k_z c),
\end{equation}
where $|{t_x}| \gg |{t_y}| \gg |{t_z}|$.
In the limit of weak interactions and low
densities these quasi-one-dimensional
systems exhibit a well defined Fermi surface which is open, being formed
of two separate sheets which intersect the Brillouin zone boundaries.
We work with the Hamiltonian
$
H = H_{kin} + H_{int},
$
where the kinetic energy and the chemical potential contribution are
$
H_{kin} = \sum_{{\bf k},\alpha} (\epsilon_{\bf k} - \mu) 
\psi_{{\bf k}, \alpha}^{\dagger} \psi_{{\bf k}, \alpha},
$
and the interaction is
\begin{equation}
\label{eqn:hint}
H_{int} = 
{1 \over 2} \sum_{{\bf k} {\bf k^{\prime}} {\bf q}} 
\sum_{\alpha \beta \gamma \delta}
V_{\alpha \beta \gamma \delta} ({\bf k}, {\bf k^{\prime}})
b_{\alpha \beta}^{\dagger} ({\bf k}, {\bf q})
b_{\gamma \delta} ({\bf k^{\prime}}, {\bf q})
\end{equation}
with
$
b_{\alpha \beta}^{\dagger} ({\bf k}, {\bf q}) = 
\psi_{-{\bf k} + {\bf q}/2, \alpha}^{\dagger}
\psi_{{\bf k} + {\bf q}/2, \beta}^{\dagger},
$
where $\alpha$, $\beta$, $\gamma$ and $\delta$ 
are spin indices and 
${\bf k}$, ${\bf k}^{\prime}$ and ${\bf q}$ represent 
linear momenta. 
In the case of weak spin-orbit coupling
and triplet pairing, the model interaction tensor can be chosen to be
$$
V_{\alpha \beta \gamma \delta} ({\bf k}, {\bf k^{\prime}})
= \Gamma_{\alpha \beta \gamma \delta} 
V_{\Gamma} ({\bf k}, {\bf k^{\prime}})
\phi_{\Gamma} ({\bf k}) \phi^{*}_{\Gamma} ({\bf k^{\prime}}),
$$
where 
$\Gamma_{\alpha \beta \gamma \delta} = {\bf v}_{\alpha \beta} \cdot
{\bf v}_{\gamma \delta}^{\dagger}/2$ with 
${\rm v}_{\alpha \beta} = (i\sigma \sigma_y)_{\alpha \beta}$. In addition,
the interaction $V_{\Gamma}$ corresponds to the irreducible 
representation $\Gamma$ with basis function
$\phi_{\Gamma} ({\bf k})$ representative of the orthorombic group. 
In the case of strong spin-orbit coupling the interaction 
$$
V_{\alpha \beta \gamma \delta} ({\bf k}, {\bf k^{\prime}})
= 
V_{\Gamma} ({\bf k}, {\bf k^{\prime}})
\left[
\Phi_{\Gamma} ({\bf k}) \cdot {\bf v}_{\alpha \beta}
\right]
\left[
\Phi^{*}_{\Gamma} ({\bf k^{\prime}}) \cdot {\bf v}_{\gamma \delta}/2
\right]
$$
where the interaction $V_{\Gamma}$ corresponds to the irreducible 
representation $\Gamma$ with basis function vector
$\Phi_{\Gamma} ({\bf k})$ representative of the orthorombic group. 

In both weak and strong spin-orbit coupling, we 
can use either the equation of motion method~\cite{mineev-99}
or the functional integration method~\cite{sdm-93,sdm-97} 
in the zero center of mass momentum pairing approximation 
(which corresponds to the BCS limit in weak coupling)
to obtain 
the anomalous Green's function
\begin{equation}
\label{eqn:anom-green}
F_{\alpha \beta} ({\bf k}, i\omega_n )  = 
{\Delta_{\alpha \beta}({\bf k}) 
\over  
\omega_n^2 + E_{\bf k}^2 },
\end{equation}
and the single particle Green's function
\begin{equation}
\label{eqn:sing-green}
G_{\alpha \beta} ({\bf k}, i\omega_n )  = 
- { i\omega_n + \xi_{\bf k}  \over \omega_n^2 + E_{\bf k}^2 }
\delta_{\alpha \beta},
\end{equation}
where $\xi_{\bf k} = \epsilon_{\bf k} - \mu$, $\mu$ is the
chemical potential,
$E_{\bf k} = \sqrt{ \xi_{\bf k}^2 + \Delta_{\bf k}^2  }$ is the 
quasiparticle excitation energy, and $\Delta_{\bf k}^2 \equiv 
{\it Tr} \left[
\tilde\Delta^{\dagger} ({\bf k}) \tilde \Delta ({\bf k})
\right]
/2$.
The matrix $\tilde \Delta ({\bf k})$ has matrix elements
$\Delta_{\alpha \beta} ({\bf k})$. The expressions for the 
single particle (Eq.~\ref{eqn:sing-green}) and for the 
anomalous (Eq.~\ref{eqn:anom-green}) Green's functions
are valid only in the unitary case where 
$\tilde\Delta^{\dagger} ({\bf k}) \tilde \Delta ({\bf k})$
is diagonal. We will not discuss here the non-unitary case.
Using the single particle and anomalous Green's functions defined
above and standard many body methods~\cite{mineev-99,sdm-93,sdm-97}
we obtain the familiar forms
\begin{equation}
\label{eqn:oparam}
\Delta_{\alpha \beta} ({\bf k}) = - \sum_{ {\bf k}^{\prime} }
V_{\beta \alpha \gamma \delta} ({\bf k}, {\bf k^{\prime}})
{ \Delta_{\gamma \delta} ({\bf k^{\prime}}) \over 2 E_{ \bf k^{\prime} } }
\tanh \left( { E_{\bf k^{\prime} } \over 2 T } \right), 
\end{equation}
\begin{equation}
\label{eqn:number}
N \equiv \sum_{ {\bf k} } n_{\bf k} = 
\sum_{ {\bf k} }  
\left[
n_{qp} ({\bf k}) + n_{qh} ({\bf k})
\right],
\end{equation}
where
$
n_{qp} ({\bf k}) =
\left(
1 + { \xi_{\bf k} / E_{\bf k} } 
\right) f ( E_{\bf k} )
$ and
$
n_{qh} ({\bf k}) =
\left(
1 - { \xi_{\bf k} / E_{\bf k} } 
\right) 
\left(
1 - f ( E_{\bf k} )
\right),
$
for the order parameter and number equations, respectively. 

%%\section{Possible symmetries for the order parameter}
%%\label{sec:symmetries}

Next we consider the allowed symmetries of the order parameter
$\Delta_{\alpha \beta}$ for an orthorhombic crystal~\cite{foot-03}, 
with a conventional symmetry normal state.
%i.e., we assume that the normal state does not break 
%the full lattice symmetry.
Here, the relevant crystallographic point 
group is ${ \rm D_{2h} }$, which has only one dimensional 
representations~\cite{tinkham-64}. This implies that
the order parameter matrix in the triplet channel,  
\begin{equation}
\label{eqn:op-matrix}
\tilde \Delta ({\bf k}) = 
i \left(
\Delta_{tr} ({\bf k}) {\bf d} ({\bf k}) \cdot {\bf \sigma} 
\right)
\sigma_y,
\end{equation}
must transform according to the one dimensional representations
of the orthorhombic point group ${\rm D_{2h}}$, under the assumption that
the order parameter does not break the crystal translational symmetry,
i.e., the order parameter is invariant under all primitive lattice
translations. 
Under the transformation ${\bf k} \to - {\bf k}$ the 
three-dimensional vector ${\bf d} ({\bf k})$ 
is antisymmetric (odd), while 
the function $\Delta_{tr} ({\bf k})$ is symmetric (even).

Here, we will be interested in triplet states which do not break time 
reversal symmetry, and our analysis will be confined to zero 
magnetic field only.
In Tables~\ref{tab:1} and~\ref{tab:2}
we summarize the group theoretical analysis 
for $\tilde \Delta ({\bf k})$, in the
weak and strong spin-orbit coupling cases respectively.
The tables include the state nomenclature, the 
vector ${\bf d} ({\bf k})$, and the type of zeros of the
quasiparticle excitation spectrum $E_{\bf k}$, when $\tilde \mu = 
\mu - {\rm min}[\epsilon_{\bf k}]$ is positive. 
In Table~\ref{tab:1}, the vector (0,0,1) is indicated up to an arbitrary 
rotation in spin space. In Table~\ref{tab:2},
the numerical coefficients A, B and C are determined
through 
Eqs.~(\ref{eqn:oparam}) and~(\ref{eqn:number}). 
It is crucial to emphasize that the basis functions
$X({\bf k}), Y({\bf k})$ and $Z ({\bf k})$ transform like 
$k_x, k_y$ and $k_z$ under the crystallographic point group operations. 
However, these functions cannot 
be chosen to be equal to $k_x, k_y$ and $k_z$
as done in the work by Lebed, Machida, and
Ozaki~\cite{lebed-00} (LMO),
since the Fermi surface $\epsilon ({\bf k}) = \mu$ intersects
the Brillouin zone along the y and z directions.
Thus, it is necessary to take into account the
periodicity of the order parameter matrix $\tilde \Delta ({\bf k})$
and of the order parameter vector ${\bf d} ({\bf k})$ in
reciprocal (momentum) space.  
As a result, the minimal basis set must be periodic and may be chosen
to be $X({\bf k}) = \sin (k_x a)$, 
$Y({\bf k}) = \sin (k_y b)$ and $Z ({\bf k}) = \sin (k_z c)$. 

For weak spin-orbit coupling (Table~\ref{tab:1}) 
the only candidate for weak attractive interaction 
($\tilde \mu > 0$) is the state $^3B_{3u} (a)$, where  
the quasiparticle excitation spectrum $E_{\bf k}$ has no zeros and 
is fully gapped. 
For strong spin-orbit coupling (Table~\ref{tab:2}) 
there are three candidates for weaker attractive interaction 
($\tilde \mu > 0$), i.e., the states $A_{1u}, B_{1u}$
and $B_{2u}$, where the quasiparticle excitation spectrum 
$E_{\bf k}$ may have no zeros and may be fully gapped. 
When $\tilde \mu > 0$, 
the state $A_{1u}$ is fully gapped only
for $A \ne 0$ and for any value of $B$ and $C$; 
the state $B_{1u}$ is fully gapped only 
for $B \ne 0$ and for any value of $A$ and $C$; and
the state $B_{2u}$ is fully gapped only  
for $C \ne 0$ and for any value of $A$ and $B$.
Note that in the case of strong attractive interactions
where $\tilde \mu < 0$, the excitation spectrum $E_{\bf k}$ is fully
gapped~\cite{duncan-00} for all states in both tables.

\begin{table}[b]
\caption{Weak spin-orbit coupling.}
%\squeezetable
\begin{tabular}{|l|c|c|}
State & ${\bf d} ({\bf k})$ & $E_{\bf k} = 0 $ $(\tilde \mu >0)$ \cr
\hline
$^3A_{1u}(a)$ & $(0,0,1)XYZ$ & lines \\ %\cr
\hline
$^3B_{1u}(a)$ & $(0,0,1)Z$  & lines \\ %\cr
\hline
$^3B_{2u}(a)$ & $(0,0,1)Y$ & lines \\ %\cr
\hline
$^3B_{3u}(a)$ & $(0,0,1)X$ & none \\ %\cr
\end{tabular}
\label{tab:1}
\end{table}
\begin{table}[b]
\caption{Strong spin-orbit coupling.}
%\squeezetable
\begin{tabular}{|l|c|c|}
State & ${\bf d} ({\bf k})$ & $E_{\bf k} = 0 $ $(\tilde \mu >0)$ \cr
\hline
$A_{1u}$ & $(AX,BY,CZ)$ & none, points or lines \cr
\hline
$B_{1u}$ & $(AY,BX,CXYZ)$ & none or lines \cr
\hline
$B_{2u}$ & $(AZ,BXYZ,CX)$ & none or lines \cr
\hline
$B_{3u}$ & $(AXYZ,BZ,CY)$ & points or lines \\
\end{tabular}
\label{tab:2}
\end{table}

Next we turn our attention to the calculation of the 
quasiparticle density of states (QDOS) 
and uniform spin susceptibility at low
temperatures for two potential triplet candidate states, which 
are fully gapped: 
a) the weak spin-orbit coupling
state $^3B_{3u} (a)$;
b) the strong spin-orbit coupling states $A_{1u}$.~\cite{foot-04}

%%\section{Quasiparticle Density of States}
%%\label{sec:qdos}

The QDOS for these different
symmetries can be obtained from 
the single particle Green's function as
\begin{equation}
{\cal N} (\omega) = - {1 \over \pi} {\it Tr} \sum_{\bf k} 
{\it Im }
G_{\alpha \beta} ({\bf k}, i\omega_n = \omega + i\delta), 
\end{equation}
where $G_{\alpha \beta} ({\bf k}, i\omega_n)$ is defined in 
Eq.~(\ref{eqn:sing-green}). 
The QDOS, shown in Fig.~1 %\ref{fig:one}%
can be measured in STM experiments.
Although these experiments have not yet been performed
in ${\rm (TMTSF)_2 ClO_4}$ and ${\rm (TMTSF)_2 PF_6}$, 
our theoretical predictions
can serve as qualitative guides for the extraction of gaps and 
symmetry dependent features when experimental results 
become available. In particular, STM measured
gaps could be compared with gaps measured either 
thermodynamically (e.g., specific heat) 
or in transport experiments (e.g., thermal conductivity~\cite{behnia-97}).
%
%%[ht!]
\vskip -0.6cm
\begin{figure}
%\vskip 4.3cm
\begin{center}
\epsfxsize=8.0cm
\epsfysize=6.0cm
%epsfysize=8.0cm
\epsfbox{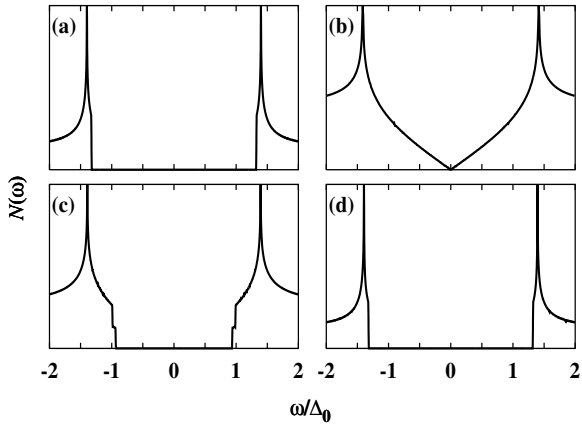}
%epsfxsize=6.0cm
%epsfysize=8.0cm
%\epsfbox{dos-strongLS.eps}
\vskip 0.2cm
\caption{Plot of QDOS
versus frequency:
the weak spin-orbit coupling state $^3B_{3u}(a)$ is shown in (a);
the strong spin-orbit coupling state $A_{1u}$ is shown
in (b) for $A = 0.01$, $B = \sqrt{2 - A^2}$, $C = 0$;
in (c) for $A = B = 1$, $C = 0$; 
in (d) for $A = \sqrt{2 - B^2}$, $B = 0.01$, $C = 0$. 
The parameters used are $|t_x| = 1000 K$, $|t_y| = 100 K$ and
$|t_z| = 5 K$, $\Delta_{tr} = \Delta_0 = 3.0 K$
and $\mu = - 250 K$. 
}
\vskip -0.4cm
\end{center}
\label{fig:one}
\end{figure}
We compare in Fig.~1 the QDOS for the states
$^3B_{3u}(a)$ (weak spin-orbit coupling) and 
$A_{1u}$ (strong spin-orbit coupling) for various values
of the constants A, B, and C. The symmetry dependent features
of the QDOS are manifested through the magnitude of the order parameter 
vector ${\bf d} ({\bf k})$. For the $^3B_{3u}(a)$ state
$|{\bf d} ({\bf k})| \propto |\sin(k_x a)|$,
while for the $A_{1u}$ state
$|{\bf d} ({\bf k})| \propto  
\sqrt{ A^2|\sin(k_x a)|^2 +  B^2|\sin(k_y b)|^2 +  C^2|\sin(k_z c)|^2}$.
In Fig.~1 
(b), (c) and (d), we study
only the case corresponding to $C = 0$, where $A \gg B$, $A = B$,
and $A \ll B$, respectively. Notice that  
Fig.~1(d) 
is nearly identical to 
Fig.~1(a) 
given that $|{\bf d} ({\bf k})|$ 
is essentially the same in this case. Furthermore, notice that
while the position of the peaks in 
Figs.~1
(a), (b), (c) and (d) are essentially the same
in scaled units ($\omega_{p}/\Delta_0 \approx \pm 1.40$), 
the corresponding gap sizes in scaled units are 
respectively $\omega_{g}/\Delta_0 = 1.32;0.01;0.94, 1.32$. 
Gap sizes, peaks and the general shape of the QDOS
should be in principle identifiable in an STM experiment. 
However, such experiments alone cannot uniquely determine the 
symmetry of the order parameter in 
triplet quasi-one-dimensional superconductors,
since the QDOS depends only on $|{\bf d} ({\bf k})|$.

%%\section{spin susceptibility}
%%\label{sec:susceptibility}

Now, we turn our attention to the calculation of the
spin susceptibility tensor, which explicitly depends on both 
the magnitude and direction of ${\bf d} ({\bf k})$,
and, thus, may help ellucidate the symmetry of the order 
parameter in quasi-one-dimensional superconductors.
The spin susceptibility tensor for a triplet superconductor is 
$$
\chi_{mn} (q_{\mu})
= -\mu_{B}^{2}
(P_{mn})_{\alpha \beta \gamma \delta}
\left[
A_{\alpha \beta \gamma \delta}(q_{\mu})
+
S_{\alpha \beta \gamma \delta}(q_{\mu})
\right],
$$
where we use the Einstein summation convention, 
and the four-vector $q_{\mu} = ({\bf q}, i\nu)$. 
The tensor $(P_{mn})_{\alpha \beta \gamma \delta} = 
(\sigma_{m})_{\alpha \beta}(\sigma_{n})_{\gamma \delta}$,
contains Pauli spin matrices, and the 
tensors
$$
A_{\alpha \beta \gamma \delta}(q_{\mu}) =
{1 \over \beta} \sum_{{\bf k}, i\omega}
F^{\dagger}_{\alpha \gamma}(-{\bf k} + {\bf q}, -i\nu + i\omega)
F_{\beta \delta}({\bf k},i\omega),
$$
$$
S_{\alpha \beta \gamma \delta}(q_{\mu}) =
{1 \over \beta} \sum_{{\bf k}, i\omega} 
G_{\delta\alpha}({\bf k} - {\bf q},-i\nu+i\omega)
G_{\beta \gamma}( {\bf k},i\omega)
\delta_{\delta \alpha}\delta_{\beta \gamma}
$$
contain the Green's functions
described in Eqs.~(\ref{eqn:anom-green}) and~({\ref{eqn:sing-green}),
and $\beta = 1/k_B T$.
For $\omega \to 0$ and ${\bf q} \to {\bf 0}$,
\begin{equation}
\label{eqn:chi-triplet}
\chi_{mn} ({\bf 0}, 0) = 
\sum_{\bf k} 
\left[\chi_{mn,1} ({\bf k})
+ \chi_{mn,2} ({\bf k})
\right],
\end{equation}
where the ${\bf k}$-dependent tensors have the forms
$$
\chi_{mn,1} ({\bf k}) = \chi_{\parallel} ({\bf k}) 
{\it Re}~{\hat d}^{*}_{m} ({\bf k}) {\hat d}_{n} ({\bf k}),
$$
$$
\chi_{mn,2} ({\bf k}) = \chi_{\perp} ({\bf k}) 
\left(
\delta_{mn} - {\it Re}~{\hat d}^{*}_{m} ({\bf k}) {\hat d}_{n} ({\bf k}) 
\right),
$$
with
${\hat d}_{n} ({\bf k}) = {d}_{n} ({\bf k})/ |d_{n} ({\bf k})| $.
The parallel component is 
\begin{equation}
\label{eqn:chi-parallel}
\chi_{\parallel} ({\bf k}) = 
- 2 \mu_B^2 { \partial f (E_{\bf k}) \over \partial E_{\bf k} }, 
\end{equation}
while the perpendicular component is  
\begin{equation}
\label{eqn:chi-perpendicular}
\chi_{\perp} ({\bf k}) = 
2 \mu_B^2 
{ d \over d\xi_{\bf k} }
\left[ {\xi_{\bf k} \over 2 E_{\bf k}}
\left( 1 - 2 f(E_{\bf k}) \right)
\right]. 
\end{equation}
This result is more general than the expression quoted 
in LMO~\cite{lebed-00} in a couple of ways. First, the expression
derived in Eqs.~(\ref{eqn:chi-triplet}),~(\ref{eqn:chi-parallel}), 
and ~(\ref{eqn:chi-perpendicular}) includes particle-hole 
symmetry effects. Second they are valid at finite T. They do not
include, however, Fermi liquid corrections.

In Fig.~2,
we show the theoretical uniform $\chi_{mn}$ 
only for the $^3 B_{3u}$ and $A_{1u}$ states~\cite{foot-04},
where triangles correspond to $\chi_{11}$
circles to $\chi_{22}$, and squares to $\chi_{33}$.
It is known experimentally (Knight
shift)~\cite{lee-00}
that the spin susceptibility of ${\rm (TMTSF)_2 PF_6}$ 
for ${\bf H} \parallel {\hat b}^{\prime}$
is very close to $\chi_N$. Experiments for magnetic field along
other directions have not yet been performed. 
Thus, for definiteness, we choose the unit vectors $\hat 3$ $(m = 3)$,
${\hat 2}$ $(m = 2)$ and ${\hat 1}$ $(m = 1)$ 
to point along the $b^{\prime}$, $a$ and $c$ direction, respectively.

For the orthorhombic symmetry $\chi_{mn}$ is diagonal, and
is calculated here under the assumption of constant
${\bf d}({\bf k})$, i.e., the direction of ${\bf d}({\bf k})$
is not changed upon application of a small magnetic field. 
In the case of state $A_{1u}$ (strong spin-orbit coupling) 
a small magnetic field cannot rotate ${\bf d}({\bf k})$
which is pinned to a particular lattice direction. 
Here, $\chi_{mn}$ is still
diagonal, however the diagonal components are not equal in general.
Thus, the experimentally measured $\chi_{mn}^{ex}$ and the
theoretically calculated $\chi_{mn}^{th}$ (at constant ${\bf d}({\bf k})$)
should agree for small enough magnetic fields.
However, in the case of state $^3B_{3u}(a)$ (weak spin-orbit coupling) 
a small magnetic field can always rotate ${\bf d}({\bf k})$
to be perpendicular to ${\bf H}$, 
and thus minimize the magnetic free-energy
$F_{mag} = - H_m \chi_{mn}H_n/2$. In this case, 
$\chi_{mn}^{ex} \approx \chi_{N} \delta_{mn}$, where
$\chi_N$ is the normal state value, for any direction ${\bf H}$.
Thus, $\chi_{mn}^{ex}$ and $\chi_{mn}^{th}$ (shown in Fig.~2)
are different.  
%
%%[ht!]
\begin{figure}
\begin{center}
\vskip -0.6cm
\epsfxsize=8.0cm
\epsfysize=6.0cm
\epsfbox{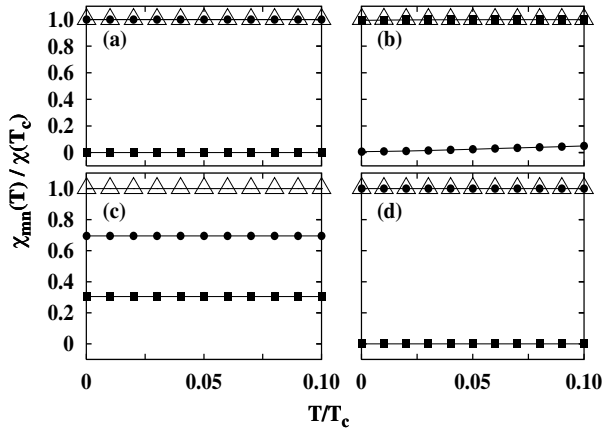}
\vskip 0.3cm
\caption{ Plot of the theoretical 
uniform spin susceptibility tensor components
$\chi_{11}$ (triangles); $\chi_{22}$ (circles);
$\chi_{33}$ (squares) at low temperatures. 
The weak spin-orbit coupling state $^3B_{3u}$ is shown in (a);
strong spin-orbit coupling state $A_{1u}$ is shown
in (b) for $A = 0.01$, $B = \sqrt{2 - A^2}$, $C = 0$;
in (c) for $A = B = 1$, $C = 0$; 
in (d) for $A = \sqrt{2 - B^2}$, $B = 0.01$, $C = 0$. 
The parameters used are $|t_x| = 1000 K$, $|t_y| = 100 K$ and
$|t_z| = 5 K$, $\Delta_{tr} = \Delta_0 = 3.0 K$
and $\mu = - 250 K$. 
}
\vskip -0.4cm
\end{center}
\label{fig:two}
\end{figure}
In conclusion, we have studied order parameter symmetry features 
in the quasiparticle
density of states and spin susceptibility tensor of orthorhombic 
quasi-one-dimensional superconductors.
We studied both the weak and strong spin-orbit coupling cases
from a group theoretical point of view at zero magnetic field. 
Based on experimental evidence, 
we would like to suggest that the weak spin-orbit coupling state
$^3B_{3u} (a)$ is the best candidate for the order parameter
symmetry for these systems since this state is:
(1) fully gapped and consistent with thermal 
conductivity measurements~\cite{behnia-97};
(2) characterized by weak spin-orbit coupling
and consistent with weak spin-orbit coupling fits of 
$T_c (H)$ for ${\rm (TMTSF)_2 PF_6}$~\cite{lee-98} at low magnetic fields;
(3) consistent with no observable 
Knight shift when ${\bf H} \parallel 
{\hat b}^{\prime}$, and predicted to have no 
observable Knight shift for any direction of ${\bf H}$.

We would like to thank the Georgia Institute of Technology, 
NSF (Grant No. DMR-9803111), and NATO (Grant No. CRG-972261)
for financial support.


\begin{references}
\bibitem{foot-04} We do not discuss other strong spin-orbit coupling 
states $B_{1u}$ and $B_{2u}$, which can be fully gapped, because of
space limitations.
\end{references}
\end{document}